# A New Approach for Correcting Noncovalent Interactions in Semiempirical Quantum Mechanical Methods. The Importance of Multiple-Orientation Sampling

*Sergio Pérez-Tabero, Berta Fernández, Enrique M. Cabaleiro-Lago, Emilio Martínez-Núñez and Saulo A. Vázquez\**

Departamento de Química Física, Facultade de Química, Universidade de Santiago de Compostela, 15782 Santiago de Compostela, Spain.

ABSTRACT

A new approach is presented to improve the performance of semiempirical quantum mechanical (SQM) methods in the description of noncovalent interactions. To show the strategy, the PM6 Hamiltonian was selected, although, in general, the procedure can be applied to other semiempirical Hamiltonians and to different methodologies. In this way, analytical corrections to PM6 were derived from fits to CCSD(T) – PM6 interaction energy differences. A set of small molecules was selected as representative of the common functional groups, and intermolecular potential energy curves were evaluated for the most relevant orientations of interacting molecular pairs. The resulting method, called PM6-FGC (from Functional Group Corrections), significantly improves the performance of PM6 and previous corrected SQM methods, and shows the importance of including a sufficient number of orientations of the interacting molecules in the reference data set in order to obtain well-balanced descriptions.

## 1. Introduction

One of the well-known problems inherent to semiempirical quantum mechanical (SQM) methods is the poor performance in describing noncovalent interactions.[1-2] Over the last years, many efforts have been devoted to improve the accuracy of SQM methods for noncovalent interactions, particularly those based on the neglect of diatomic differential overlap (NDDO) approximation.[3-4] The most common strategy used to ameliorate the performance of SQM methods in calculations of intermolecular interactions has been the inclusion of empirical corrections.[5-21]

Řezáč, Hobza, and their co-workers developed several generations of corrections for dispersion,[6-7, 9] hydrogen bond[7, 9] and halogen bond[8] interactions, and parameterized them within the PM6 method[22] as well as for other SQM methods. Contributions to this series of generations were also made by Korth[10] and Jensen and co-workers.[11] The final version of this series of corrections is called D3H4X, in reference to the third generation dispersion correction, fourth generation hydrogen-bonding correction, and halogen-bonding correction. In this version, the dispersion correction is the D3 proposed by Grimme et al. for Density Functional Theory (DFT),[23] but without including the $1/r^8$ term, which was considered to yield no significant improvement in the case of SQM methods.[9] For these methods, Řezáč and Hobza found a specific error in the description of interactions between hydrocarbons, namely, the overestimation of interaction energies and the underestimation of equilibrium distances.[9] To solve this problem, they included a repulsive term for all pairs of hydrogen atoms. The function used to improve the description of hydrogen bonding includes a polynomial function of degree 7 in the donor-acceptor distance, which is scaled by an angular term (dependent on the acceptor-hydrogen-donor angle) and a proton transfer term that varies with the hydrogen position. If the system contains charged groups, an additional factor is included to increase the strength of the correction. Finally, the correction used for halogen bonding consists of an exponential term.[8] The D3H4X correction, as well as other generations of corrections, have been implemented in the MOPAC2016 program.[24]

The procedure adopted by Řezáč and Hobza to parameterize the D3H4 corrections was the following.[9] First, they fitted the hydrogen-bonding correction, including the contribution from dispersion in the calculated energies of the considered hydrogen-bonded complexes. For the fittings, they performed least-squares optimizations, minimizing the root-mean-square error of the interaction energy when compared to reference data obtained at the Coupled Cluster Singles and Doubles with perturbative Triples correction/Complete Basis Set (CCSD(T)/CBS) level of calculation. Specifically, as benchmark set, they used the S66 database,[25-27] which includes dissociation curves for 66 noncovalent complexes that exhibit dispersion, hydrogen bonds, and mixed dispersion/electrostatic interactions.

Truhlar, Gao, and their co-workers developed the polarized molecular orbital (PMO) method,[12-16] based on a NDDO Hamiltonian that includes polarization functions on hydrogen atoms. In addition, to improve the description of dispersion interactions, they added the first damped dispersion term developed by Grimme.[28-29] This dispersion correction was previously used by Hillier and co-workers[5, 20] in conjunction with the AM1[30] and the PM3[31] Hamiltonians. The final versions of the PMO method, that is, PMO2[15] and PMO2a,[16] have been found to accurately describe polarization effects as well as noncovalent complexation energies. The PMO2 method was parametrized for all compounds containing H, C and O atoms, and the PMO2a version is an extension of PMO2 to new functionalities, which includes parameters for amino nitrogen groups and molecules containing sulfur–oxygen bonds. Parameterizations of the PMO Hamiltonians were carried out using a genetic algorithm, which has the advantage of efficiently explore the search space to find near optimal solutions when the number of fitting parameters is large. The above PMO versions have been implemented on the MOPAC 5.022mn package.[32]

The work of Thiel and co-workers directed to improve the reliability of SQM methods also deserves some attention. They developed the ortogonalization-corrected methods OM$x$[33-35] and ODM$x$,[17] which include significant improvements in the semiempirical Hamiltonian, thus leading, in general, to better results in comparison with NDDO-based methods that make use of the modified neglect of diatomic overlap (e.g., AM1 or PM6). These semiempirical Hamiltonians needed to incorporate dispersion corrections to improve the description of noncovalent interactions. In particular, they include Grimme's D3 dispersion correction[23, 36] with the Becke-Johnson damping function,[37-39] as well as Axilrod-Teller-Muto three-body terms,[23, 40] which ameliorate the description of large dense systems.[41-42] For the recent ODM$x$ methods, several training sets were used, including the abovementioned S66 data set.[25-27] Parameterization of semiempirical Hamiltonians and correction potentials for noncovalent interactions is a key issue, and the procedure followed within the ODM$x$ methods is extensively discussed in the recent article by Dral et al.[17]

The above studies led to remarkable improvements in SQM methods for the evaluation of noncovalent interactions. In general, the corrections for dispersion and hydrogen bonding interactions are modelled by potential functions based on physically sound formulas. In addition, the training sets used for parameterizations are quite large, which may ensure a wide range of applicability. However, and as it will be shown later in the present work, errors in the description of noncovalent interactions may be significant, depending on the relative orientation of the interacting molecules. This can be a consequence of possible shortcomings in popular data sets, which in general only include the most relevant configurations of interacting molecules.

In this paper, we present, as a proof-of-concept study, an alternative way to develop analytical corrections for SQM methods to improve the description of noncovalent interactions. The idea is based on previous chemical dynamics studies in which pairwise intermolecular potentials were parameterized through fittings to a series of intermolecular potential energy curves (IPECs) that emphasize the different atom-pair potentials exhibited by the interacting molecules.[43-47] Following the strategy used to develop potentials for interactions of peptides with self-assembled monolayers of perfluorinated alkanes,[47] we selected small molecules as representatives for typical functional groups, and evaluated IPECs for all possible molecular pairings. Specifically, we have chosen methane, formic acid, and ammonia, which give six different pair combinations: the three dimers and the $CH_4/HCO_2H$, $CH_4/NH_3$, and $NH_3/HCO_2H$ pairs of molecules. We developed empirical corrections for the PM6 method, which can be used not only for interactions between hydrocarbons, carboxylic acids and amines, but also for other compounds, as will be justified later. Actually, as validation dataset, we considered a collection of different conformers of the diglycine and dialanine dimers, obtained through automated exploration of the corresponding potential energy surfaces (PESs). The novelty of our approach is the introduction of two important features. First, and most important, the inclusion of several orientations of the interacting molecules in the database, which is crucial to obtain well-balanced corrections. And second, the use of general corrections to take into account that SQM methods have significant limitations not only to accurately describe dispersion interactions but also electrostatics, induction and exchange repulsion.

## 2. Methods

Intermolecular potential energy curves for the six pairs of molecules indicated above were calculated using CCSD(T),[48] and the augmented correlation consistent polarized valence triple-zeta basis set aug-cc-pVTZ.[49] For test purposes, the IPECs were additionally evaluated employing DFT with the B3LYP functional,[50-52] including the D3 dispersion correction with the Becke-Johnson damping scheme,[37-39] and with the valence triple-zeta polarization def2-TZVP basis set.[53] The IPECs were computed using the supermolecular approach with frozen intramolecular geometries and correcting the interaction energy for basis set superposition error (BSSE) through the counterpoise method.[54-55] The intramolecular geometries were obtained by B3LYP-D3/def2-TZVP optimizations. Several orientations of the interacting molecules were selected to stress the different pair-type interactions. Specifically, for each pair of molecules, the number of orientations was at least equal to the number of the different pair-type interactions. A proper selection of orientations is crucial to obtain well-balanced corrections. These electronic structure calculations were performed with the ORCA 4.0 program and the default frozen core approximation.[56-57]

The general expression of the noncovalent potential-energy correction developed in this work for the PM6 method is written as a pairwise sum of the form:

$$E_{\text{corr}} = \sum_i \sum_j f_{\text{cut}}(r_{ij}) \times \left\{ A_{ij} e^{-B_{ij} r_{ij}} + \frac{C_{ij}}{r_{ij}^{n_{ij}}} \right\} \quad (1)$$

where indexes $i$ and $j$ refer to atoms belonging to different interacting molecules, and $r_{ij}$ is the interatomic distance between atoms $i$ and $j$. Parameters $A_{ij}$, $B_{ij}$ and $C_{ij}$ (real numbers) as well as $n_{ij}$ (integers) depend on the nature of the considered pair of atoms. $f_{\text{cut}}(r_{ij})$ is a cutoff function introduced to remove the correction at very short $r_{ij}$ distances:

$$f_{\text{cut}}(r_{ij}) = \left(1 + \tanh\left(s_{ij}(r_{ij} - d_{ij})\right)\right)/2 \tag{2}$$

where $s_{ij}$ is a parameter that controls the strength of the damping for the interaction between atoms $i$ and $j$, and $d_{ij}$ is the distance at which the cutoff function takes the value ½. The $n_{ij}$ parameters were not fixed to 6; rather, they were allowed to vary around this value. Also, the $A_{ij}$ and $C_{ij}$ parameters may be either positive or negative. We notice that eq 1 should be regarded as a practical correction, without any physical interpretation. However, one may expect the functional form given by eq 1, based on Buckingham's potential,[58] to work reasonably well because this potential can model intermolecular interactions with pretty good accuracy.

The above parameters were obtained through fittings to differences between the interaction energies calculated at the reference level (i.e., CCSD(T), or B3LYP-D3 for test purposes) and those computed with the PM6 method. The SQM calculations were carried out with the MOPAC2016 program.[24] We used a least-squares nonlinear fitting procedure based on a genetic algorithm, as implemented in our GAFit code,[59] and with the following objective function, $\chi^2$:

$$\chi^2(\mathbf{a}) = \sum_{i=1}^{N}[y_i - f(x_i; \mathbf{a})]^2 \times w_i \tag{3}$$

where $(x_i, y_i)$ represents one of the $N$ data points, $\mathbf{a}$ is the collective variable formed by the total number of fitting parameters and $f(x_i; \mathbf{a})$ is the value of the model function at $x_i$ (i.e., a particular geometry of the interacting molecules). The square of the difference between $y_i$ (i.e., a CCSD(T)–PM6 energy difference) and the corresponding model value, calculated with eqs 1 and 2, may be multiplied by a weighting factor ($w_i$) assigned to each data point. In this work, we obtained corrections for the PM6 method to reproduce CCSD(T)/aug-cc-pVTZ as well as B3LYP-D3/def2-TZVP interaction energies (the latter for test purposes).

To validate our model function and parameterization strategy, we used a dataset formed by a collection of different conformations of the diglycine and dialanine dimers, obtained by an automated exploration of the PESs of these dimers at the PM6-D3H4 level, using the AutoMeKin package,[60-62] which has an interface with the MOPAC2016 program.[24] Although AutoMeKin has mainly been designed to discover and simulate chemical reaction mechanisms, it includes an option for obtaining stationary points for intermolecular complexes.[63] The benchmark level for this validation was B3LYP-D3/def2-TZVP, correcting the interaction energies for BSSE.

**3. Results and discussion**

**3.1 Formic acid dimer.** The formic acid molecule has five chemically non-equivalent atoms, that is, all the atoms are non-equivalent. Therefore, for this system, there are 15 different types of pairwise interactions. Consequently, for the fittings, we included 15 orientations that emphasize the distinct pairwise interactions, as well as an additional orientation corresponding to the global minimum of the formic acid dimer, which exhibits double hydrogen bonding. These 16 orientations are depicted in Figure 1.

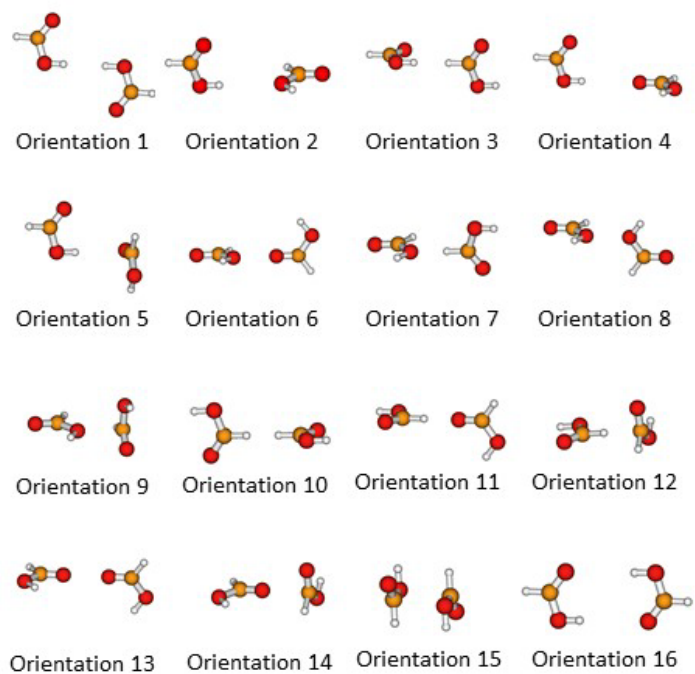

**Figure 1.** Orientations of the formic acid dimer considered for the fitting.

As pointed out in the previous section, for the selected orientations we calculated intermolecular potential energy curves with the reference method, that is, CCSD(T)/aug-cc-pVTZ, and with the PM6 method, in order to develop corrections for the latter using the functional form specified in eqs 1 and 2. We used the differences between the CCSD(T)/aug-cc-pVTZ and the corresponding PM6 interaction energies as the data for the parameterizations. These energy differences have the typical forms displayed in Figure 2. In these four plots, $r$ corresponds to the distance between attacking atoms, that is, the two carboxylic hydrogens in orientation 1, for example. The form of the energy difference as a function of $r$ for this orientation resembles a typical repulsive potential, indicating that PM6 has a less repulsive IPEC than that of the reference method. By contrast, for orientation 4 the form of the CCSD(T) – PM6 plot behaves as a decaying exponential with negative amplitude ($A_{ij}$ in equation 1), thus pointing out a stronger repulsion character of the PM6 interaction potential. Most of the orientations show a well followed by a pronounced increase of the energy difference as the distance between the attacking atoms becomes shorter, as can be seen for orientation 6 (carbonyl oxygen - hydroxyl oxygen attack). The presence of a well in these plots does not mean that the CCSD(T) and PM6 IPECs exhibit potential minima (although in most cases they do). Actually, for orientation 6 the CCSD(T) and PM6 IPECs are repulsive in nature, as can be seen in Figure 3. In very few cases (only one for the formic acid dimer), the CCSD(T) − PM6 energy differences display a more complex form, showing both a minimum and a maximum, as for orientation 12, where the hydrogen attached to the carbonyl carbon attacks the carbon atom of the other molecule. Although it could be expected that equation 1 would be valid as a practical and simple correction for SQM methods, these plots further justify its use.

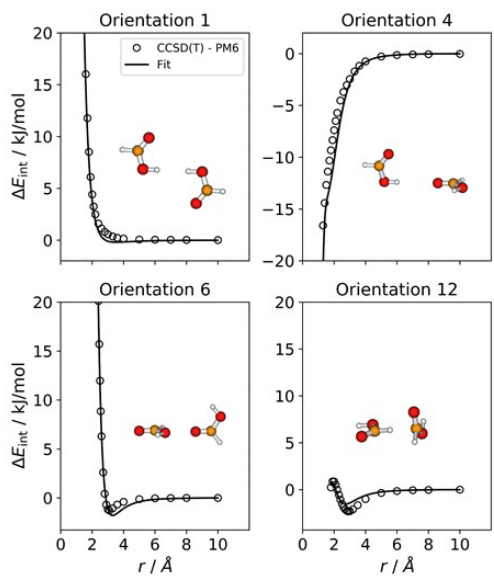

**Figure 2.** CCSD(T) − PM6 interaction energy differences (open circles) for selected orientations of the formic acid dimer. The black lines correspond to the fit (see text).

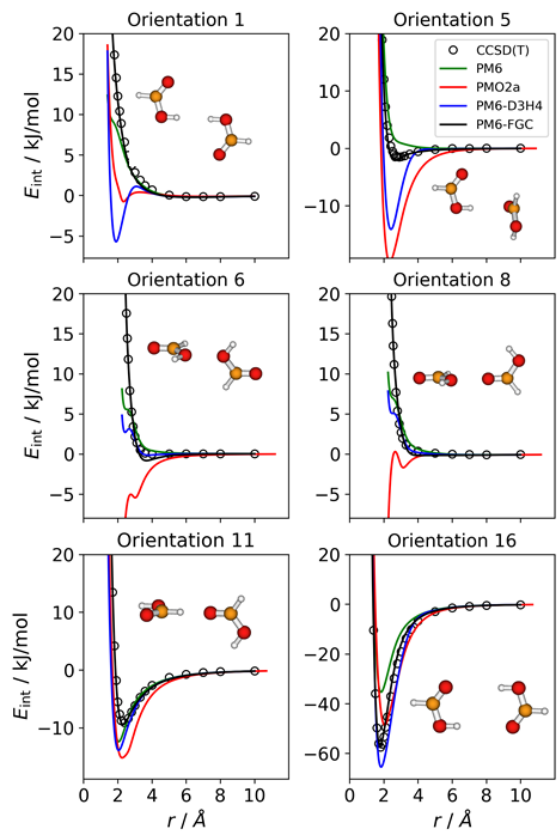

**Figure 3.** Comparison of IPECs for six selected orientations of the formic acid dimer.

Using GAFit[59] and the geometries and energy differences corresponding to the 16 orientations of the formic acid dimer, we simultaneously fitted the parameters involved in equations 1 and 2. There is not a universal, objective way to conduct a parametrization and, furthermore, with the use of genetic algorithms, one may obtain many solutions than can be equally valid. For general discussions on parameterizations for SQM methods, the reader may consult the studies on the development of the PMO2a[16] and ODM$x$[17] methods. For the formic acid dimer, since there are 15 different types of pairwise interactions, the total number of parameters is 60, without including those associated with the cutoff function given by eq 2. All the 60 parameters were fitted simultaneously. We found that including the cutoff parameters into the parameter spectrum explored by the genetic algorithm did not improve the fittings significantly. For this reason, after some analyses and to avoid overparameterization, we have chosen a value of 10 for all the $s_{ij}$ parameters, and different values for the $d_{ij}$ parameters, depending on the nature of atoms $i$ and $j$. The parameters obtained from our best fit are collected in Table 1, and the fit results are depicted in Figure 2 for some selected orientations. For simplicity, for parameters *A*, *B* and *C*, we only show two decimals; the high precision parameters are included in Table S1 in the Supporting Information (SI). Notice that we have defined atom types in much the same way as in molecular mechanics force fields. The symbols chosen in this work are shown in Figure 4.

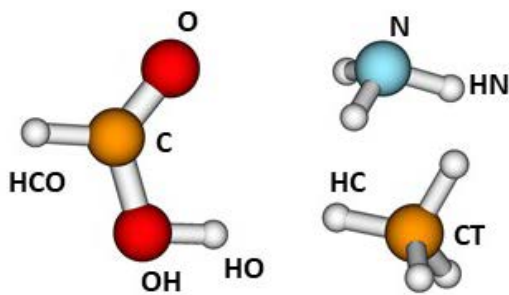

**Figure 4.** Atom types defined for ammonia, formic acid, and methane.

**Table 1.** Parameters obtained for the formic acid dimer. [a]

| Atom pair | $A$ | $B$ | $C$ | $n$ | $d$ |
|---|---|---|---|---|---|
| C–C | 42893.32 | 5.46 | −3366.81 | 6 | 1.8 |
| C–O | 27995.90 | 3.39 | −44.35 | 8 | 1.7 |
| C–OH | 109395.04 | 3.84 | −1457.06 | 7 | 1.7 |
| C–HO | −139930.20 | 5.10 | 414.89 | 6 | 1.2 |
| C–HCO | −140000.00 | 5.22 | 320.24 | 6 | 1.2 |
| O–O | 259958.92 | 3.97 | −582.59 | 5 | 1.7 |
| O–OH | 264408.19 | 3.84 | −1500.00 | 6 | 1.7 |
| O–HO | 11854.89 | 3.72 | −549.08 | 5 | 1.0 |
| O–HCO | 13519.65 | 3.49 | −1258.70 | 8 | 1.2 |
| OH–OH | 237572.26 | 3.70 | −1002.71 | 5 | 1.7 |
| OH–HO | −1470.19 | 2.77 | −50.79 | 6 | 1.0 |
| OH–HCO | 4830.25 | 2.97 | −176.34 | 6 | 1.2 |
| HO–HO | 1000.00 | 3.00 | 93.62 | 4 | 1.0 |
| HO–HCO | 27144.19 | 4.32 | −641.46 | 8 | 1.0 |
| HCO–HCO | 7722.72 | 3.47 | −184.79 | 5 | 1.0 |

[a] The units are such that the potential energy is in kJ/mol and distances in Å.

Adding to the PM6 interaction potential the corrections calculated with equations 1 and 2, and the parameters fitted in this work, results in our PM6-FGC method. We have chosen the FGC acronym, from functional group corrections, to emphasize the idea of specific parameters being obtained for different functional groups. However, as discussed later in this paper, the parameters determined with our selected molecules may be transferable to other related functional groups, at least as a first approximation. Actually, in many force fields, similar atom types share common parameters for nonbonded interaction terms.

Figure 3 compares the reference IPECs with the PM6-FGC interaction curves for six selected orientations of the formic acid dimer. The global minimum corresponds to orientation 16. The IPECs for the remaining orientations are displayed in Figure S1 in the SI. For comparison, we include the PM6 curves as well as those obtained with the PM6-D3H4[9] and PMO2a[16] methods, which are implemented in the freely distributed MOPAC2016[24] and MOPAC 5.022mn[32] programs, respectively. It would be interesting to include results of calculations using the ODM*x* method;[17] however, to our knowledge, the code in which this method is implemented is not freely available.[64] As can be seen from Figures 3 and S2, the PM6-FGC curves (black lines) agree well with the CCSD(T)/aug-cc-pVTZ data (black open circles).

The IPECs calculated with the PMO2a method (red lines) display remarkable discrepancies with the reference curves. Strikingly, for orientations 6 and 8, as well as for orientation 13 (see SI), this method shows an unphysical behavior, since the interaction energy in the repulsive region decreases as the distance between the attacking atoms diminishes. These orientations correspond to configurations that emphasize the interaction between oxygen atoms. Clearly, a revision of this method is required to improve its performance. For this reason, for the remaining systems under investigation here, we have not considered the PMO2a method any further.

The PM6-D3H4 potential energy curves are displayed as blue lines in the figures. The D3H4 corrections were developed using a training set based on CCSD(T)/CBS data, so that slight deviations may be expected because our IPECs were obtained with the aug-cc-pVTZ basis set. However, for orientation 1, which corresponds to the attack between carboxyl hydrogens, the PM6-D3H4 method exhibits a clear minimum. This orientation is predicted to be repulsive at the CCSD(T)/aug-cc-pVTZ level. Also, for orientation 5, the PM6-D3H4 method gives a significant minimum, which contrasts with the small well depth predicted by the reference calculations. It is also worth mentioning that, for orientations 6 and 8, PM6-D3H4 and PM6 exhibit an unphysical behavior in the repulsive region (around 5 kJ/mol). Although the S66x8 data set[25] employed by Hobza and co-workers comprises a wide range of complexes, and the D3H4 corrections are able to describe noncovalent interactions for the most relevant orientations of interacting molecules, our results point out some deficiencies in these corrections, which may be especially problematic for dynamics studies, where all orientations may be sampled. The source of these deficiencies comes from an important drawback of the S66x8 database, namely the fact that, in general, it only includes the most relevant orientation for each selected complex. For carboxylic acids, this database includes the acetic acid dimer in its most attractive orientation, i.e., that exhibiting a double hydrogen bond (the equivalent of orientation 16 for the formic acid dimer, Figure 1).

**3.2 Ammonia dimer.** This dimer shows three different types of pairwise interactions, and therefore we need at least three different orientations. In this work, we considered the four orientations depicted in Figure 5, which compares the IPECs obtained with PM6 (green lines), PM6-D3H4 (blue lines), and PM6-FGC (black lines), with those determined with the benchmark method (open circles). As can be seen from Figure 5, the PM6-D3H4 method exhibits substantial deficiencies, similar to those encountered in the formic acid dimer. Specifically, it predicts a remarkable minimum for orientation 1 (H···H attack), which is clearly repulsive at the reference level. In addition, for orientation 2, which exhibits hydrogen bonding, the PM6-D3H4 method clearly overestimates the strength of the interaction.

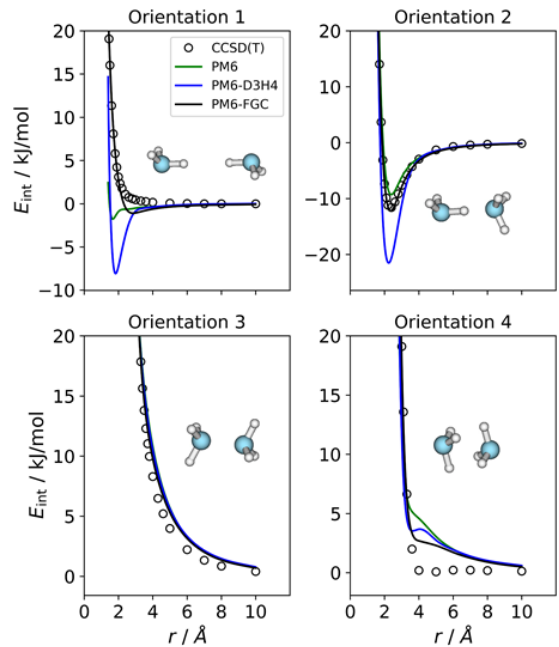

**Figure 5.** Comparison of IPECs for the considered orientations of the ammonia dimer.

Among the systems considered in this study, the ammonia dimer was the most challenging. Actually, using our simple expression for the analytical correction (eqs 1 and 2), we were not successful at obtaining a good fit, as reflected in Figure 5. Particularly, for orientation 4, PM6-FGC, as well as PM6 and PM6-D3H4, show curves more repulsive than that predicted with the benchmark method. One way to improve the fit is to add a pseudoatom, to model the effect of the nitrogen lone pair, as it is done in several force fields, but for this proof-of-concept presentation we wanted to keep the correction scheme as simple as possible. The parameters obtained from the ammonia dimer fit are collected in Table S1.

**3.3 Methane dimer.** The IPECs evaluated for this dimer are displayed in Figure 6, which also describes the orientations selected in this work. In principle, this system appears to be the simplest one among those studied here. As can be seen, both PM6-D3H4 and PM6-FGC, using the parameters shown in Table S1, exhibit IPECs in satisfactory agreement with the CCSD(T)/aug-cc-pVTZ curves. The underestimation of the dispersion interaction in the PM6 method is clear, but the worse performance predicted with this method appears for orientation 1 (i.e., H⋯H attack), which shows a significant minimum at a quite short H⋯H distance.

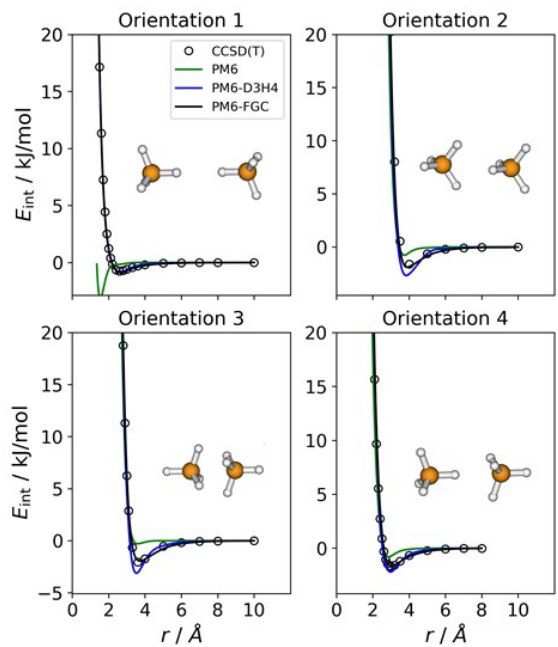

**Figure 6.** Comparison of IPECs for the considered orientations of the methane dimer.

**3.4 Ammonia-formic acid complex.** For this complex we considered ten orientations, that is, the same number as that of the different types of pairwise interactions. The best fit for this system led to the parameters reported in Table S1. Figure 7 depicts four selected orientations, together with their IPECs. The plots for the remaining orientations are shown in Figure S2. The most attractive orientation exhibits a hydrogen bond between the carboxylic hydrogen and the ammonia nitrogen (orientation 2). Both the PM6-D3H4 (blue line) and the PM6-FGC (black line) methods satisfactorily describe the interaction for this orientation. However, for several other orientations (1, 5, 7, 9 and 10), the PM6-D3H4 method predicts minima with potential well depths larger than those obtained through CCSD(T)/aug-cc-pVTZ calculations.

For orientation 1, that is, the attack between carboxylic and ammonia hydrogens. the behavior of the PM6-D3H4 curve resembles that found for orientation 1 in the formic acid and ammonia dimers. In these three cases, the reference IPECs clearly exhibit repulsive character. The PM6-FGC curve shows a small deviation from the reference IPEC, similar to that exhibited in the ammonia dimer. For orientations 9 and 10, which correspond to the attack of the carbonyl carbon to ammonia hydrogen and nitrogen, respectively, the PM6-D3H4 (and PM6) curves also show remarkable discrepancies with respect to the CCSD(T) curves. For orientation 9, the PM6-FGC curve exhibits a small deviation from the benchmark IPEC. Overall (see also Figure S2), the PM6-FGC method gives a satisfactory description of the noncovalent interaction in the ammonia-formic acid complex.

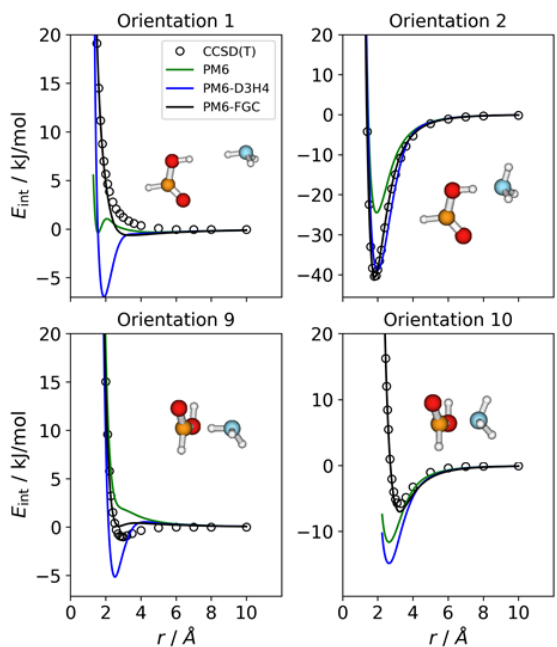

**Figure 7.** Comparison of IPECs for four selected orientations of the HCOOH/$NH_3$ complex.

**3.5 Methane-formic acid complex.** For this complex we considered ten orientations, and four of them are depicted in Figure 8, together with their corresponding IPECs. The IPECs of the remaining orientations are shown in Figure S3, and the parameters obtained in the fit are displayed in Table S1. As can be seen, the PM6-FGC curves agree well with the corresponding benchmark IPECs. The PM6-D3H4 method shows, in general, satisfactory performance, although for several orientations (e.g., 5 and 10) it predicts more attractive interactions. As expected, the PM6 interaction energies are very inaccurate, and for orientations 1 and 3 (attacks of hydrogen atoms) the corresponding IPECs display remarkable minima.

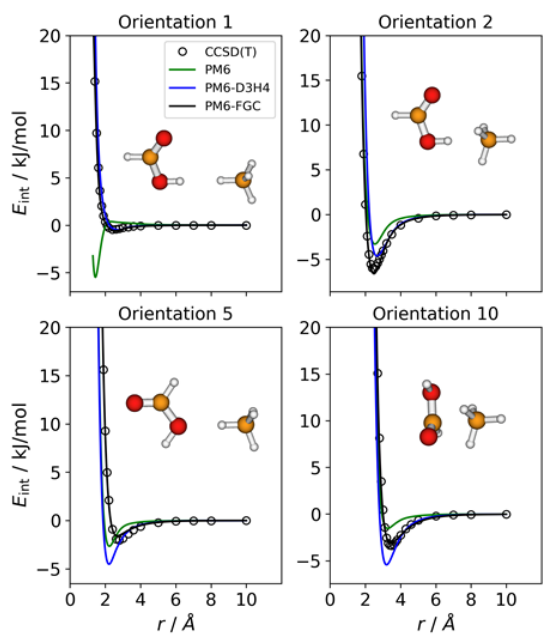

**Figure 8.** Comparison of IPECs for four selected orientations of the HCOOH/CH$_4$ complex.

**3.6 Ammonia-methane complex.** Four different orientations were considered for this complex, and they are displayed in Figure 9, together with their IPECs. The agreement between the IPECs obtained with the PM6-FGC method and those evaluated at the CCSD(T)/aug-cc-pVTZ level reflects the good quality of the fit. The PM6-D3H4 method also predicts satisfactory interaction energies, except for orientation 4, for which it provides a significant potential well depth.

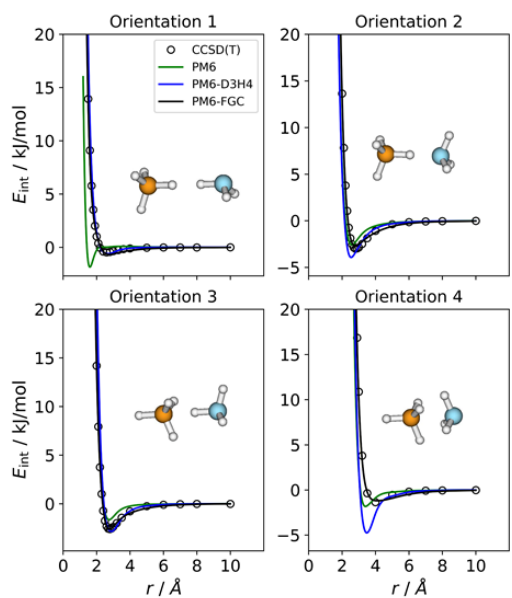

**Figure 9.** Comparison of IPECs for the considered orientations of the $CH_4/NH_3$ complex.

**3.7 Validation of the parameterization strategy.** As mentioned in Section 2, to validate our strategy we performed parameterizations using IPECs evaluated at the B3LYP-D3/def2-TZVP level. The purpose was to make a comparison of interaction energies in larger complexes (diglycine and dialanine dimers), using this DFT level as benchmark. The parameters obtained from the best fits for the six complexes considered in this paper are shown in Table S2 in the SI. The IPECs calculated with the PM6-FGC method, using the parameters reported in Table S2, are compared with the DFT curves in Figures S4 - S10. For completeness, those figures also include the CCSD(T) interaction energy curves. In general, the agreement between the PM6-FGC and the DFT curves resembles that exhibited between the IPECs obtained with our SQM method and with the CCSD(T) method. Interestingly, the B3LYP-D3/def2-TZVP curves agree very well with those determined at the CCSD(T)/aug-cc-pVTZ level; the differences between corresponding IPECs are smaller than the errors of the fits.

In principle, our corrections may be used to calculate interaction energies for systems composed of the functional groups exhibited in Figure 4. However, we additionally have applied them to compute interaction energies for conformers of the diglycine and the dialanine dimers. Although the present corrections can be expected to work well for amines and carboxylic acids, we also expect them to be satisfactory for amides (at least as a first approximation), taking into account our previous experience on the development of interaction potentials for soft-landing of protonated peptides on perfluorinated self-assembled monolayers.[47] In particular, it was found that the IPECs obtained for a Ne atom attacking the H and N atoms of ammonia are very similar to those determined for neon facing the H and N atoms of the amide group in formamide. Moreover, analytical potentials obtained from fits involving the $CF_4/NH_3$ and the $CF_4/HCOOH$ systems satisfactorily reproduced ab initio IPECs calculated for the $CF_4/HCONH_2$ complex (where the MP2/aug-cc-pVTZ level was used as benchmark).[47]

In this work, conformers of diglycine and dialanine dimers were found by automated exploration of their potential energy surfaces at the PM6-D3H4 level, using AutoMeKin,[60-63] which has an interface with the MOPAC2016 program.[24] These searches involved changes in both intramolecular and intermolecular conformations, and led to 77 and 90 different conformers for diglycine and dialanine dimers, respectively. Figure 10 shows linear correlations between the interaction energies calculated at the B3LYP-D3/def2-TZVP level and those evaluated with the SQM methods considered here, for the complexes of the diglycine dimer. For the PM6-FCG method, we include the results obtained using the parameters of Table S2 (red open circles), i.e., those to model B3LYP-D3 interaction energies, as well as the results achieved with the parameters of Table S1 (black open circles), even though the latter were developed from fits to CCSD(T) − PM6 interaction energy differences. The diagonal straight lines represent the case of perfect correlation. As can be seen, our corrections significantly improve the performance of the PM6 Hamiltonian (shown as green open circles in Figure 10). As shown in Table 2, the mean absolute errors (MAE) and mean bias errors (MBE) calculated for both PM6-FGC parameter sets are substantially smaller than those computed for PM6. The bias error is calculated as the mean of the differences between the reference values and the SQM values. In the case of the diglycine dimer, the MAE calculated for PM6-D3H4 is very similar to that of PM6 (~18 kJ/mol). Likewise, the bias values for these two methods are very similar to each other in absolute value. However, the PM6 method underestimates the strength of the interactions (negative MBE), especially for the most attractive conformers of the complexes, whereas the PM6-D3H4 method exhibits overestimation (positive MBE), especially for conformers showing from low to moderate interaction strengths.

**Table 2.** Statistical parameters[a] of the linear correlations between B3LYP-D3/def2-TZVP interaction energies and the different PM6 SQM interaction energies calculated for the conformers of the diglycine and dialanine dimers.

|                     | Diglycine dimer |       | Dialanine dimer |       |
|---------------------|-----------------|-------|-----------------|-------|
|                     | MAE             | MBE   | MAE             | MBE   |
| PM6                 | 17.9            | −15.0 | 14.3            | −14.0 |
| PM6-D3H4            | 18.6            | 15.5  | 20.3            | 19.5  |
| PM6-FGC (B3LYP-D3)  | 6.0             | 3.2   | 8.7             | 8.7   |
| PM6-FGC (CCSD(T))   | 6.4             | −4.8  | 4.6             | −0.5  |

[a] MAE and MBE values are given in kJ/mol.

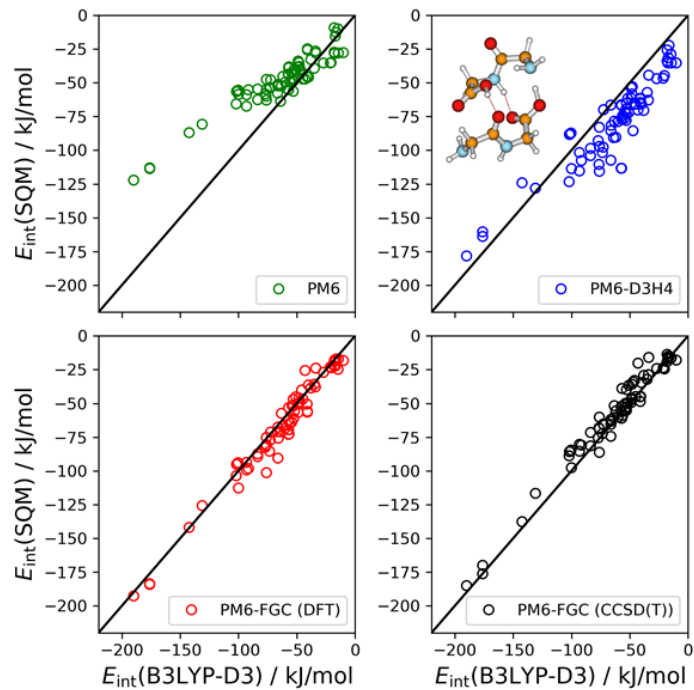

**Figure 10.** Linear correlations between the B3LYP-D3/def2-TZVP interaction energies and those calculated with PM6, PM6-D3H4, and PM6-FCG, for conformers of the diglycine dimer. The molecular drawing corresponds to the conformer with the largest interaction strength as calculated at the DFT level.

The linear correlations evaluated for the conformers of the dialanine dimer are depicted in Figure 11. As in the case of the diglycine dimer, the improvement obtained with both sets of FGC parameters is remarkable, especially with those developed to reproduce CCSD(T) interaction energies. As already noticed, the B3LYP-D3 and the CCSD(T) interaction energies calculated for the molecular pairings considered in this study are very similar to each other, the differences being, in general, smaller that the errors of the fits (Figures S4 - S10). Therefore, it is not a surprise that both sets of FGC parameters led to good correlations. The MAE and MBE values (Table 2) calculated for PM6-D3H4 are somewhat larger than those computed for PM6.

Again, the PM6 and the PM6-D3H4 methods lead to underestimation and overestimation, respectively, of the interaction energies. Although a direct comparison with PM6-D3H4 results cannot be made because the latter corrections were determined from a CCSD(T)/CBS reference, the results of the present work provide evidence of deficiencies of the PM6-D3H4 method. These deficiencies are mainly the result of using, for the parameterization scheme, a data set (i.e., the S66 database[25, 27]) that does not include sufficient orientations of the interacting molecules.

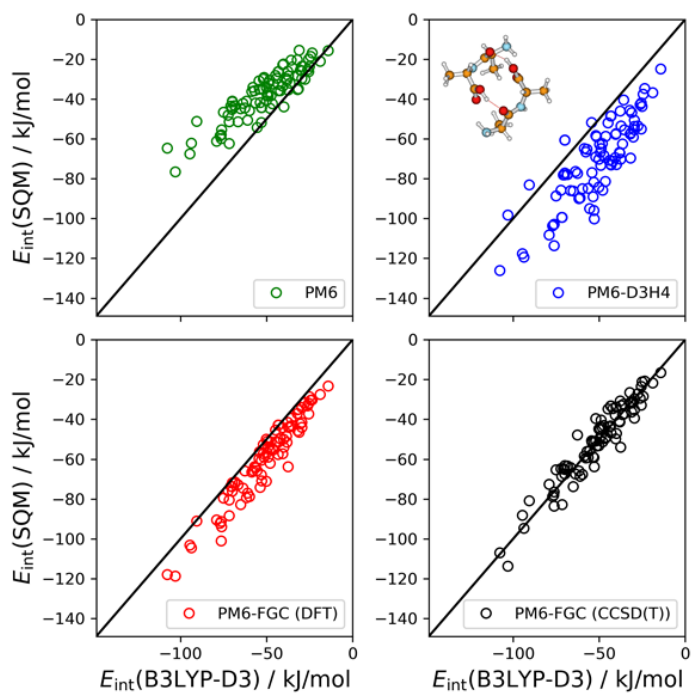

**Figure 11.** Linear correlations between the B3LYP-D3/def2-TZVP interaction energies and those calculated with PM6, PM6-D3H4, and PM6-FCG, for conformers of the dialanine dimer. The molecular drawing corresponds to the conformer having the largest interaction strength as calculated at the DFT level.

## 4. Conclusions

We have presented a new strategy, i.e. the PM6-Functional Group Corrections (PM6-FGC) method, to develop analytical corrections for semiempirical quantum mechanical methods, aimed at improving the description of noncovalent interactions. For this proof-of-concept presentation, we have selected the PM6 SQM method. Employing this Hamiltonian and the CCSD(T)/aug-cc-pVTZ level for benchmarking, we calculated intermolecular potential energy curves for several orientations of pairs of small molecules, which are selected as representatives of different functional groups. Specifically, we have considered ammonia, formic acid, and methane, which result in a total of six molecular pairings. A simple mathematical expression, which should be considered as a practical way, without any physical meaning, is used to represent the corrections to improve the performance of SQM methods. The parameters of the analytical corrections were evaluated from fits to differences between interaction energies calculated at the reference level and those evaluated at the SQM level.

In general, the IPECs obtained with the method parameterized in this work are in good agreement with those determined at the CCSD(T)/aug-cc-pVTZ level, and considerably improve those provided by successfully corrected SQM methods. In this way, this work emphasizes the importance of including, in the databases, sufficient orientations of the interacting molecules; a fact that is crucial to develop well-balanced corrections. In addition, the validation test performed on the diglycine and dialanine dimers supports our parameterization strategy.

Although the method described here involves simple pairwise functions, more accurate corrections could be developed by using or adding alternative functions, as well as by introducing three-body terms of the type of the Axilrod-Teller-Muto potential.[23, 40] In future work, we will extend our parameterizations to other types of functional groups. A Python script to compute PM6-FGC corrections is available upon request to the authors.

ASSOCIATED CONTENT

Parameters obtained from fits to CCSD(T)/aug-cc-pVTZ – PM6 interaction energy differences (Table S1); comparisons of IPECs not shown in the manuscript for the formic acid dimer (Figure S1), HCOOH/$NH_3$ (Figure S2), and the HCOOH/$CH_4$ complex (Figure S3); parameters obtained from fits to B3LYP-D3/def2-TZVP – PM6 interaction energy differences (Table S2); comparison of IPECs for PM6-FGC using the parameters of Table S2 (Figures S4 – S10).

AUTHOR INFORMATION

**Corresponding Author**

Saulo A. Vázquez

*E-mail: saulo.vazquez@usc.es

ORCID

Sergio Pérez-Tabero: 0000-0002-9744-1798

Berta Fernández: 0000-0001-6686-6534

Enrique M. Cabaleiro-Lago: 0000-0001-5848-6523

Emilio Martínez-Núñez: 0000-0001-6221-4977

Saulo A. Vázquez: 0000-0002-2473-4557

**Notes**

The authors declare no competing financial interest.

ACKNOWLEDGMENT

The authors thank Ministerio de Ciencia e Innovación for financial support (Grant # PID2019-107307RB-I00).

*Supporting Information for*

A New Approach for Correcting Noncovalent Interactions in Semiempirical Quantum Mechanical Methods. The Importance of Multiple-Orientation Sampling

*Sergio Pérez-Tabero, Berta Fernández, Enrique M. Cabaleiro-Lago, Emilio Martínez-Núñez and Saulo A. Vázquez\**

Departamento de Química Física, Facultade de Química, Universidade de Santiago de Compostela, 15782 Santiago de Compostela, Spain.

**Table S1.** Parameters[a] obtained in this study from fits to CCSD(T) − PM6 interaction energy differences.

| Atom pair | $A$ | $B$ | $C$ | $n$ | $d$ |
|---|---|---|---|---|---|
| C–C | 42893.3190472424 | 5.4598807102 | −3366.8121271375 | 6 | 1.8 |
| C–O | 27995.9042602000 | 3.3948970718 | −44.3521125473 | 8 | 1.7 |
| C–OH | 109395.0353833116 | 3.8369380257 | −1457.0637041345 | 7 | 1.7 |
| C–HO | −139930.2038496929 | 5.1020748189 | 414.8938197992 | 6 | 1.2 |
| C–HCO | −140000.0000000000 | 5.2183644292 | 320.2417032341 | 6 | 1.2 |
| O–O | 259958.9226082212 | 3.9680676533 | −582.5942388369 | 5 | 1.7 |
| O–OH | 264408.1875403374 | 3.8394567506 | −1500.0000000000 | 6 | 1.7 |
| O–HO | 11854.8935349742 | 3.7222127792 | −549.0751273874 | 5 | 1.0 |
| O–HCO | 13519.6481269705 | 3.4884575095 | −1258.6965474622 | 8 | 1.2 |
| OH–OH | 237572.2581183096 | 3.6957710297 | −1002.7148647053 | 5 | 1.7 |
| OH–HO | −1470.1873319635 | 2.7709172853 | −50.7887286521 | 6 | 1.0 |
| OH–HCO | 4830.2526675351 | 2.9674920122 | −176.3412436563 | 6 | 1.2 |
| HO–HO | 1000.0000000000 | 3.0000000000 | 93.6212400214 | 4 | 1.0 |
| HO–HCO | 27144.1911543580 | 4.3231825968 | −641.4554865539 | 8 | 1.0 |
| HCO–HCO | 7722.7199722160 | 3.4714707559 | −184.7863731238 | 5 | 1.0 |
| N–N | 61875.4055004137 | 2.9904862834 | −3612.5012524276 | 6 | 1.8 |
| HN–N | 10989.4433898952 | 5.0000000000 | −488.7426002143 | 6 | 1.2 |
| HN–HN | 4315.2468073077 | 2.9989283155 | −303.6785721258 | 6 | 1.2 |
| CT–CT | 14679.1298570751 | 2.6122774689 | −8697.7844046754 | 6 | 1.8 |
| CT–HC | 2759.4788526214 | 2.3331704988 | −3365.2145725902 | 7 | 1.2 |
| HC–HC | 23191.5310551067 | 4.1669519728 | −438.6065288928 | 7 | 1.2 |
| C–N | 4332.7389816339 | 2.1868168324 | −1198.8773165181 | 6 | 1.8 |
| C–HN | −19662.8192413641 | 4.3472002705 | −218.2573539966 | 5 | 1.2 |
| O–N | 267199.2992337945 | 3.7072609993 | −1807.5754419054 | 6 | 1.8 |
| O–HN | 7705.6093568323 | 3.3584229348 | −992.3029658675 | 7 | 1.2 |
| OH–N | 45953.2913034582 | 2.9088230491 | −994.7334746403 | 6 | 1.8 |
| OH–HN | 13909.5661329071 | 4.3000411033 | −587.1231801738 | 7 | 1.2 |
| HO–N | 223771.5619445240 | 4.8719915879 | −2469.9103571769 | 6 | 1.2 |
| HO–HN | 2532.6943750437 | 2.7928019233 | −107.1511489081 | 6 | 1.2 |
| HCO–N | 7148.5500670664 | 3.4041284907 | −702.8188886440 | 9 | 1.2 |
| HCO–HN | 12662.6808302311 | 4.1268636910 | −248.7198520746 | 8 | 1.2 |

[a] The units are such that the potential energy is in kJ/mol and distances in Å.

**Table S2.** Parameters[a] obtained in this study from fits to B3LYP-D3 − PM6 interaction energy differences.

| Atom pair | *A* | *B* | *C* | *n* | *d* |
|---|---|---|---|---|---|
| C–C | 87222.6781703499 | 4.5192960449 | −2614.2390222503 | 6 | 1.8 |
| C–O | 18423.1251143120 | 3.2165256155 | −4.7565336414 | 10 | 1.7 |
| C–OH | 101988.8220580146 | 3.7620617158 | −801.8979286559 | 6 | 1.7 |
| C–HO | −118682.0300621045 | 5.3124974616 | 49.9211435603 | 4 | 1.2 |
| C–HCO | −91515.9156668177 | 5.3134822945 | 5.6811250600 | 2 | 1.2 |
| O–O | 157533.3119267652 | 3.8363490696 | −249.6964065606 | 9 | 1.7 |
| O–OH | 229576.1122743190 | 3.7398102896 | −1494.5123518379 | 6 | 1.7 |
| O–HO | −12321.2844572512 | 5.1464744095 | −106.3638810776 | 3 | 1.0 |
| O–HCO | 7911.4730614547 | 3.4463958590 | −735.0028283044 | 9 | 1.2 |
| OH–OH | 282168.5215761239 | 3.7405298249 | −1073.4903717942 | 5 | 1.7 |
| OH–HO | −3746.8891857739 | 3.4364552152 | −49.8028148968 | 4 | 1.0 |
| OH–HCO | 3900.4353604300 | 2.9066422385 | −380.8650480007 | 8 | 1.2 |
| HO–HO | 2029.7914153477 | 3.0685212894 | 27.0433227950 | 2 | 1.0 |
| HO–HCO | 10653.6465827195 | 4.2574799189 | −396.7704915296 | 11 | 1.0 |
| HCO–HCO | 9870.5095997493 | 3.4861656955 | −289.7235420067 | 5 | 1.0 |
| N–N | 14325.7433949299 | 2.3474982183 | 0.0 | 0 | 1.8 |
| HN–N | 67180.6938579479 | 4.0291906191 | −2542.4108143032 | 6 | 1.2 |
| HN–HN | 2669.0457913742 | 2.6179842324 | −301.4938231690 | 6 | 1.2 |
| CT–CT | 15857.1701602370 | 2.6980129877 | −770.4222566934 | 5 | 1.8 |
| CT–HC | 1146.3278506151 | 2.2146125958 | −1170.1797569593 | 6 | 1.2 |
| HC–HC | 18261.8993616442 | 4.3048851482 | −522.7261176189 | 10 | 1.2 |
| C–N | 8824.6374645503 | 2.7387802474 | 40.0451859169 | 4 | 1.8 |
| C–HN | −9896.4961081155 | 4.3777416739 | −283.1284987128 | 5 | 1.2 |
| O–N | 189991.8017198780 | 3.6348674855 | −0.0066170660 | 8 | 1.8 |
| O–HN | 11946.3431019849 | 4.2125802231 | −782.9420790391 | 8 | 1.2 |
| OH–N | 10971.8258111343 | 2.3045860697 | −1499.8948318794 | 6 | 1.8 |
| OH–HN | 30103.3944495938 | 5.0405312142 | −313.4604773768 | 6 | 1.2 |
| HO–N | 294065.3724309836 | 4.5821919300 | −4264.6523403123 | 6 | 1.2 |
| HO–HN | 1989.0390506399 | 2.7000114780 | −114.0461092908 | 8 | 1.2 |
| HCO–N | 8856.4806583822 | 3.7125803660 | −527.1736982525 | 10 | 1.2 |
| HCO–HN | 4914.7612479035 | 3.5584587531 | −108.0011048550 | 6 | 1.2 |

[a] The units are such that the potential energy is in kJ/mol and distances in Å.

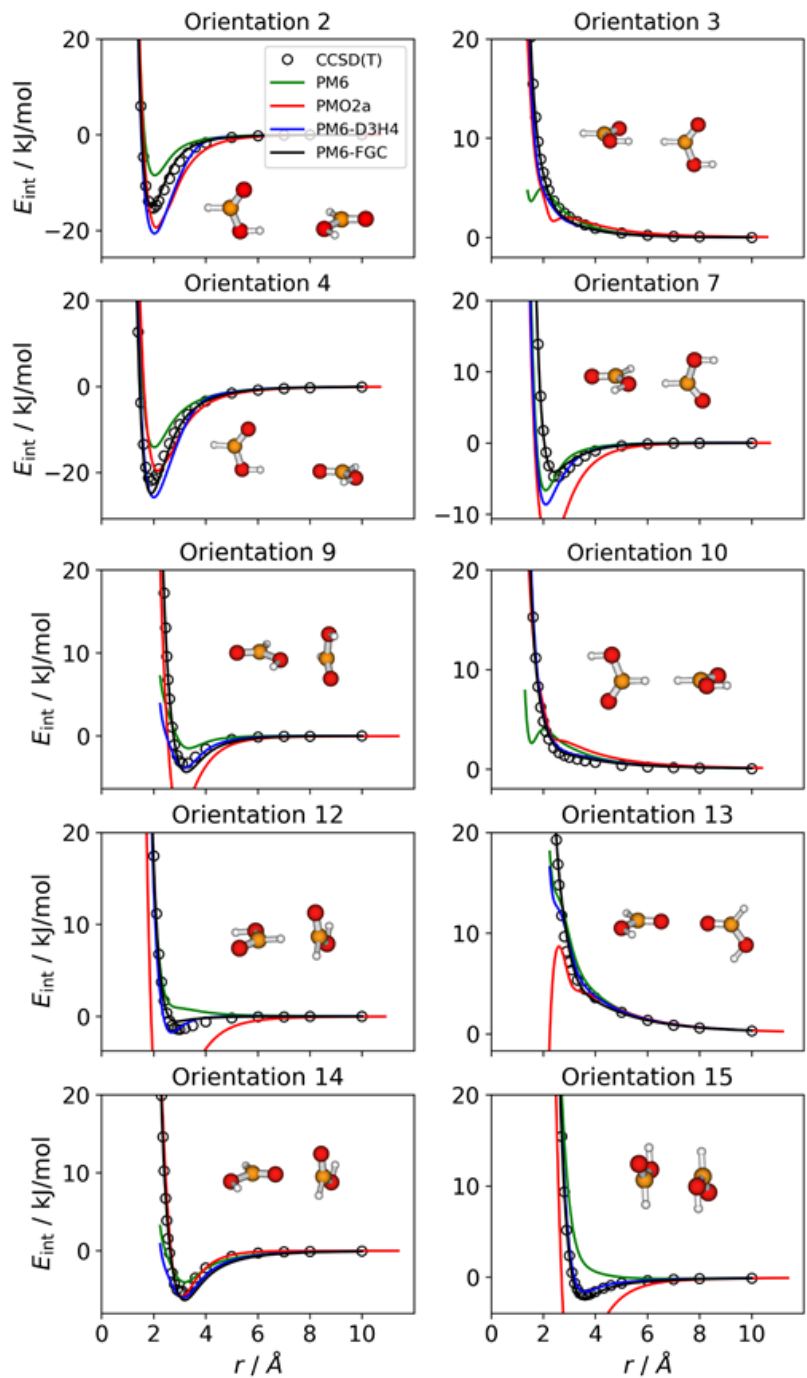

**Figure S1.** Comparison of IPECs for the orientations of the formic acid dimer not shown in the manuscript.

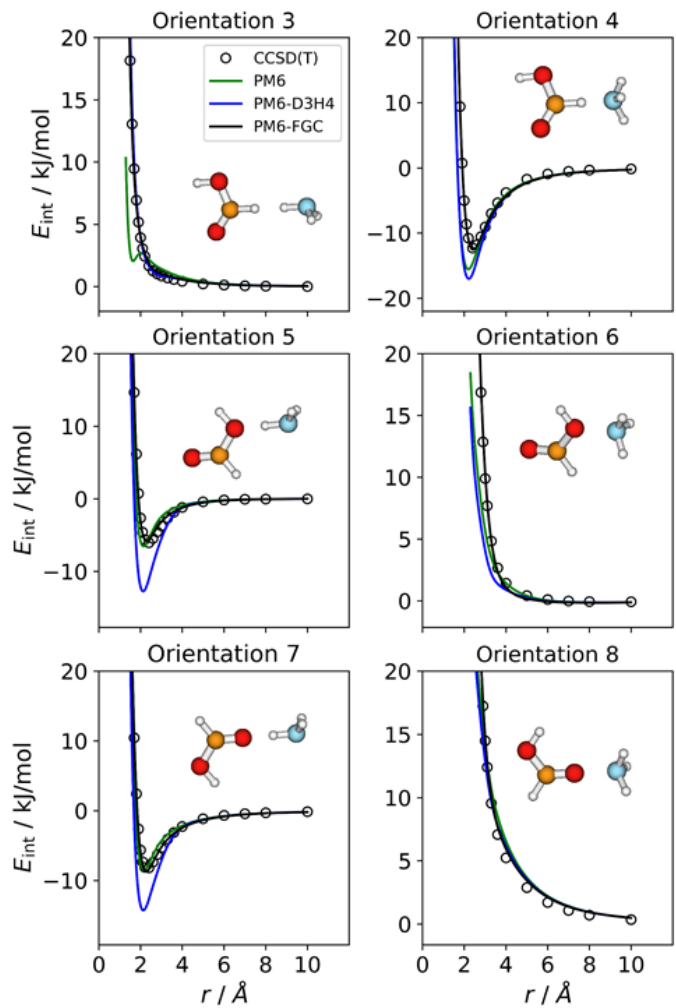

**Figure S2.** Comparison of IPECs for the orientations of the HCOOH/$NH_3$ complex not shown in the manuscript.

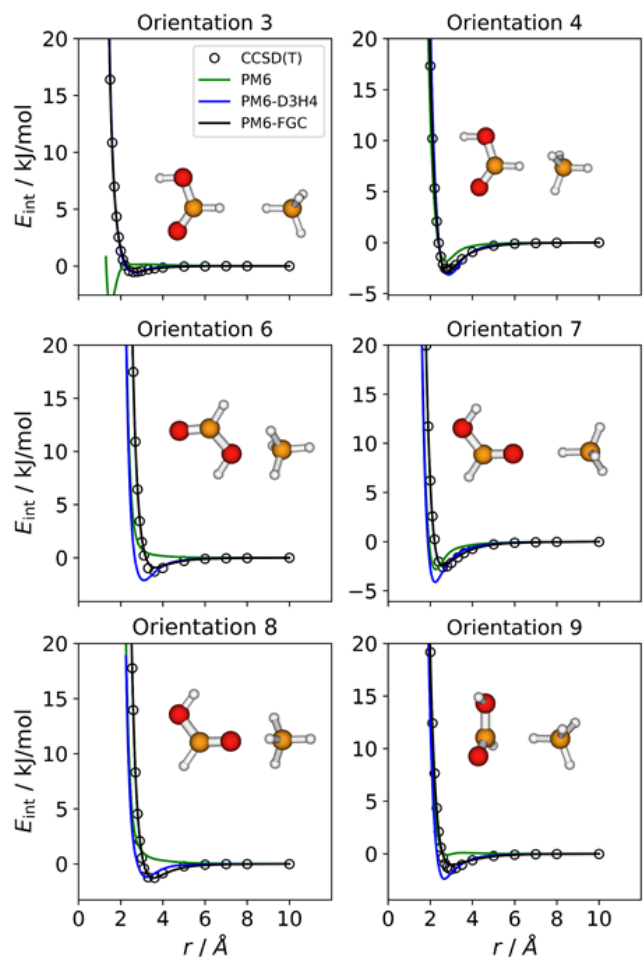

**Figure S3.** Comparison of IPECs for the orientations of the HCOOH/CH$_4$ complex not shown in the manuscript.

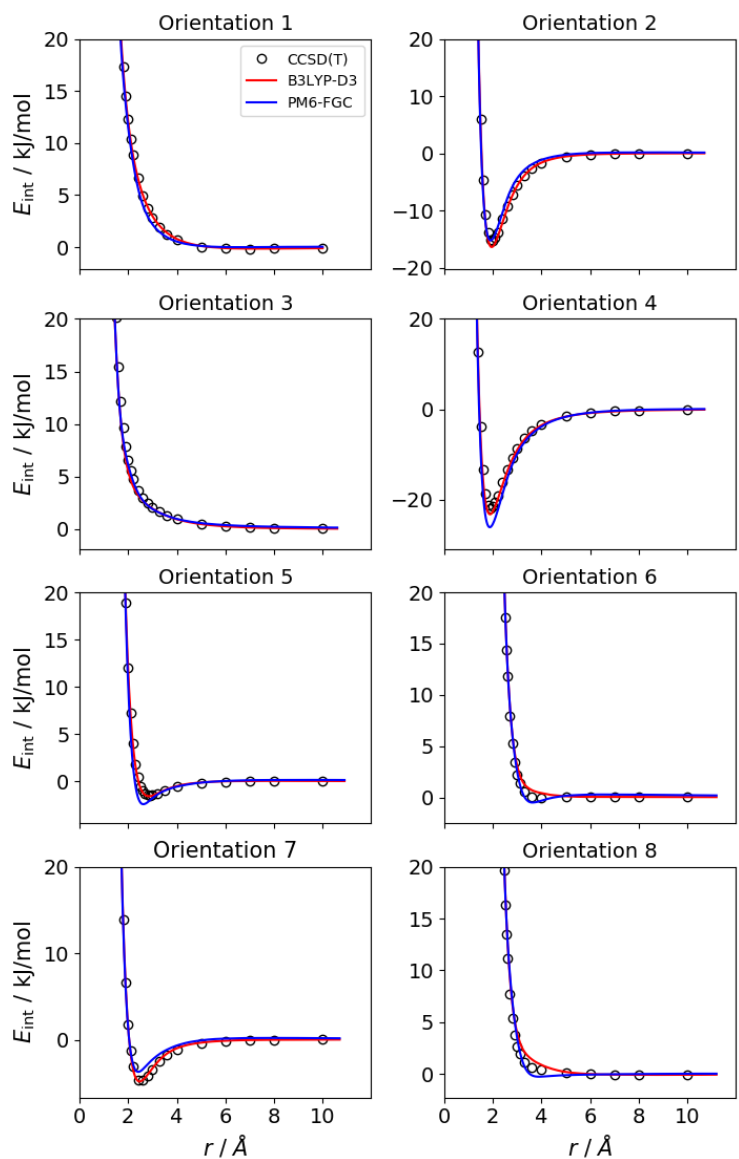

**Figure S4.** Comparison of IPECs for orientations 1-8 of the formic acid dimer. Corrections to PM6 were obtained from a fit to B3LYP-D3 – PM6 interaction-energy differences.

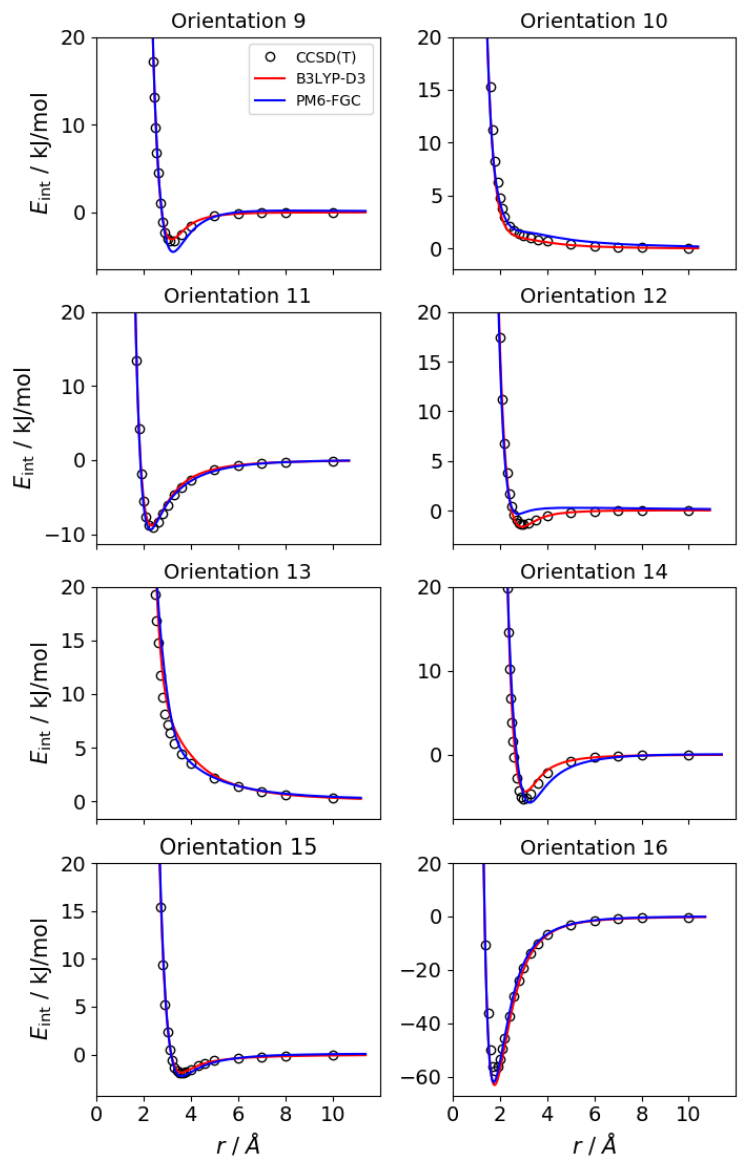

**Figure S5.** Comparison of IPECs for orientations 9-16 of the formic acid dimer. Corrections to PM6 were obtained from a fit to B3LYP-D3 – PM6 interaction-energy differences.

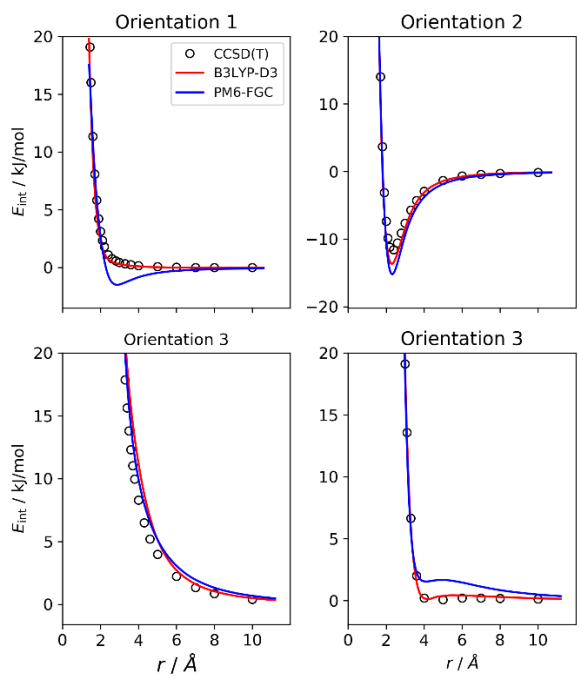

**Figure S6.** Comparison of IPECs for the orientations of the ammonia dimer. Corrections to PM6 were obtained from a fit to B3LYP-D3 – PM6 interaction-energy differences.

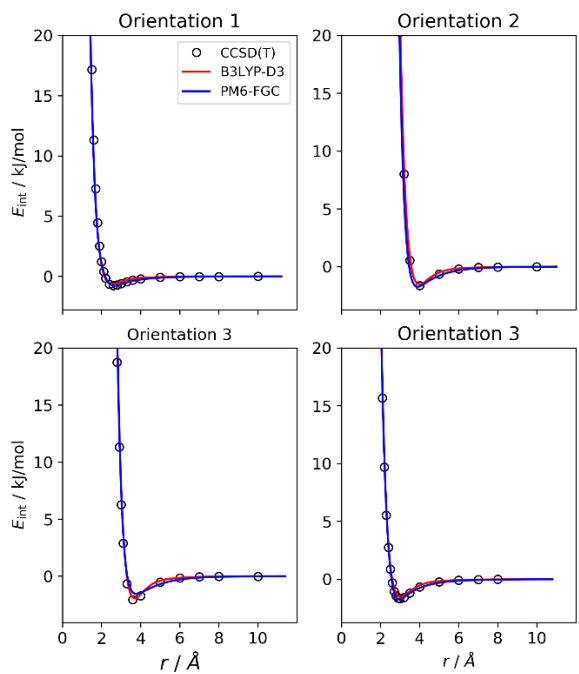

**Figure S7.** Comparison of IPECs for the orientations of the methane dimer. Corrections to PM6 were obtained from a fit to B3LYP-D3 – PM6 interaction-energy differences.

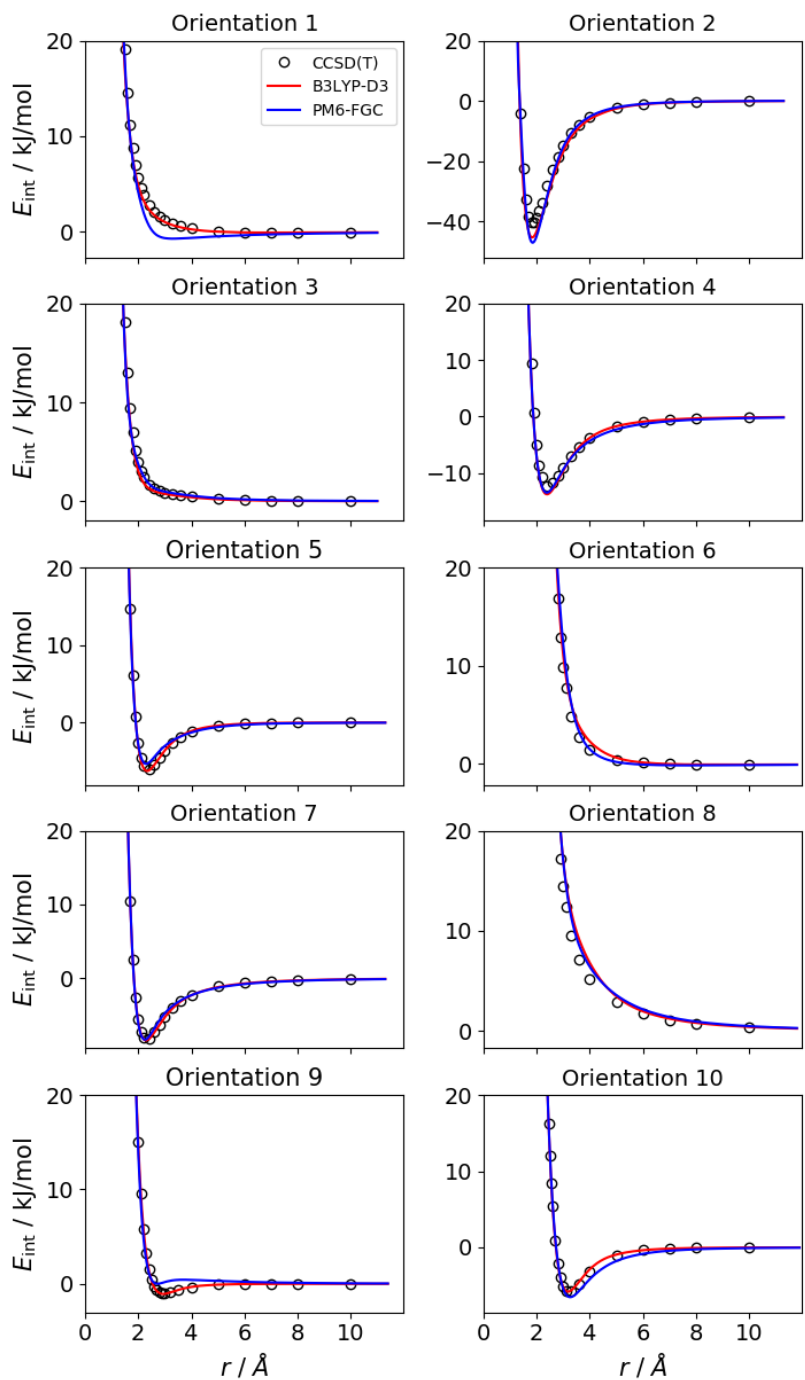

**Figure S8.** Comparison of IPECs for the orientations of the HCOOH/$NH_3$ complex. Corrections to PM6 were obtained from a fit to B3LYP-D3 – PM6 interaction-energy differences.

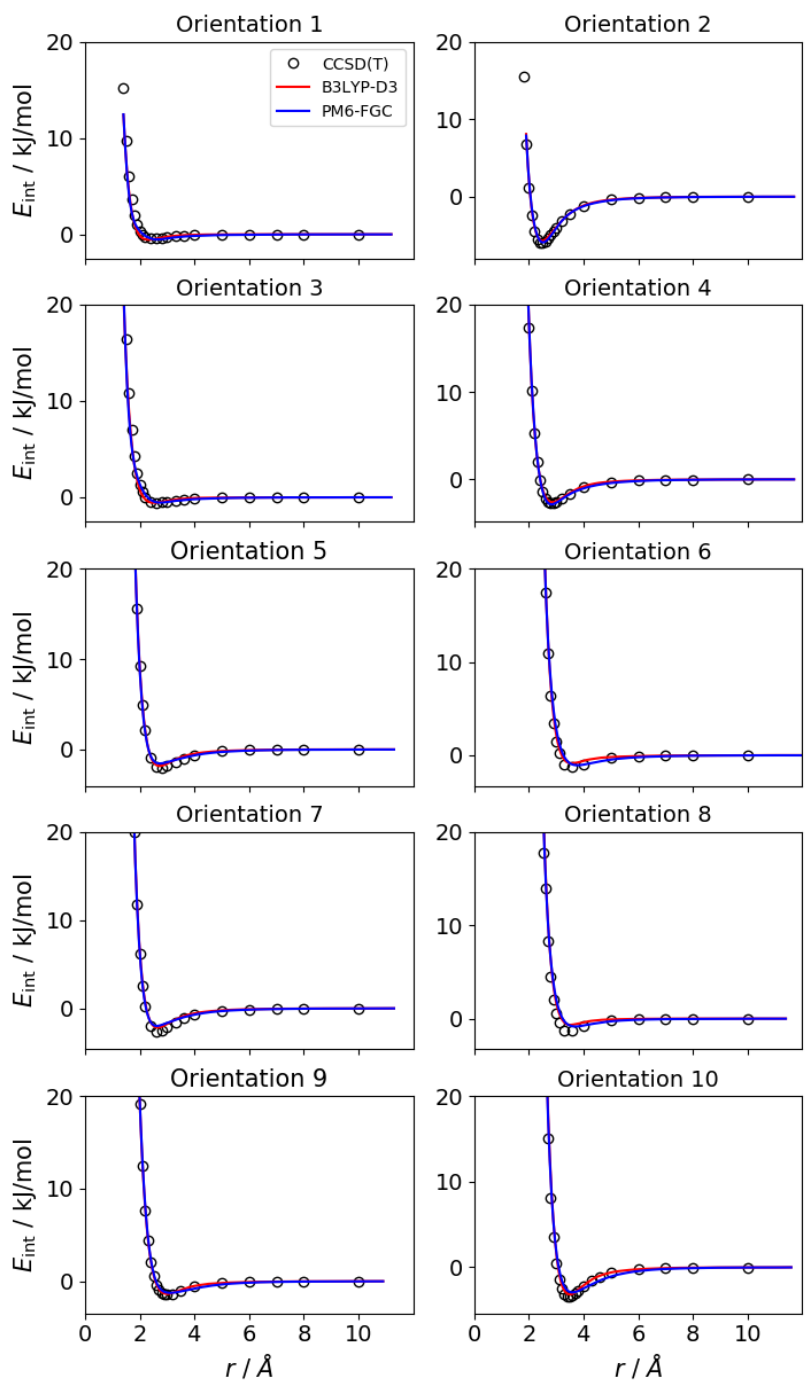

**Figure S9.** Comparison of IPECs for the orientations of the HCOOH/CH$_4$ complex. Corrections to PM6 were obtained from a fit to B3LYP-D3 – PM6 interaction-energy differences.

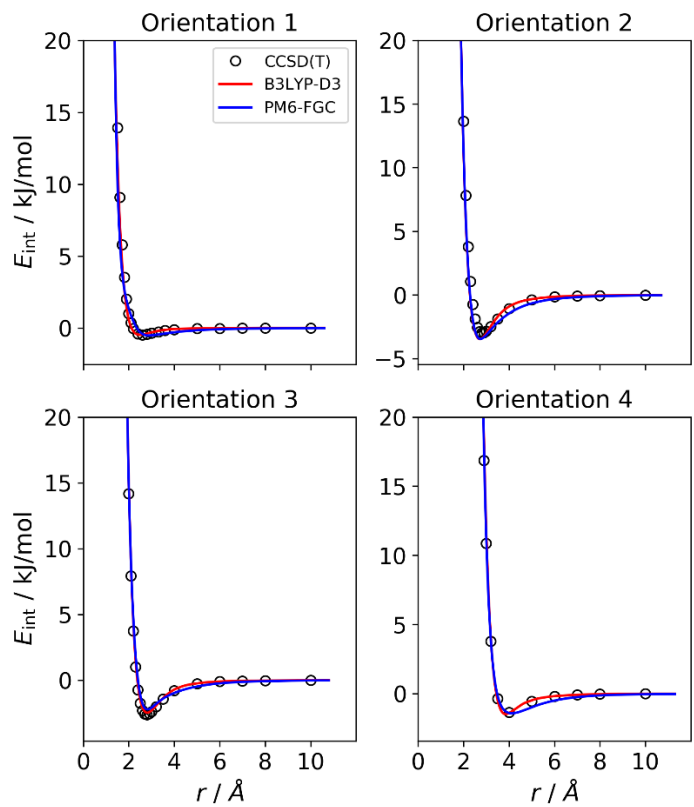

**Figure S10.** Comparison of IPECs for the orientations of the $NH_3/CH_4$ complex. Corrections to PM6 were obtained from a fit to B3LYP-D3 – PM6 interaction-energy differences.